\newif\ifAnon\Anonfalse
\newif\ifDraft\Draftfalse
\newif\ifFinal\Finalfalse

\ifFinal\Anonfalse\Draftfalse\fi
\documentclass[runningheads]{llncs}
\usepackage[T1]{fontenc}

\usepackage{hyperref}
\usepackage{cispa_defaults}
\usepackage{graphicx}
\usepackage{makecell}

\newcommand{\parhead}[1]{\textbf{#1}.\ }
\newcommand{\code}[1]{\texttt{#1}\xspace}

\newcommand{\meltdownus}{Meltdown-US-L1\xspace}

\newcommand{\IDTAttack}{LeakIDT\xspace}

\newcommand{\LabTwelveCPU}{Intel Xeon E3-1505M v5\xspace}
\newcommand{\LabTwelveOS}{Ubuntu 20.04\xspace}
\newcommand{\LabTwelveKernel}{Linux kernel 5.4.0\xspace}

\newcommand{\TSXLaptopCPU}{Intel Core i7-6600U\xspace}
\newcommand{\TSXLaptopOS}{Ubuntu 20.04\xspace}
\newcommand{\TSXLaptopKernel}{Linux kernel 5.4.0\xspace}

\begin{document}
\title{Indirect Meltdown: Building Novel Side-Channel Attacks from Transient-Execution Attacks}
\titlerunning{Indirect Meltdown}

\author{Daniel Weber \and
Fabian Thomas \and
Lukas Gerlach \and\\
Ruiyi Zhang \and
Michael Schwarz}

\authorrunning{Weber et al.}

\institute{CISPA Helmholtz Center for Information Security\\
Saarbr\"ucken, Saarland, Germany\\
\email{<firstname>.<lastname>@cispa.de}}

\maketitle

\begin{abstract}
The transient-execution attack Meltdown leaks sensitive information by transiently accessing inaccessible data during out-of-order execution.
Although Meltdown is fixed in hardware for recent CPU generations, most currently-deployed CPUs have to rely on software mitigations, such as KPTI.
Still, Meltdown is considered non-exploitable on current systems.

In this paper, we show that adding another layer of indirection to Meltdown transforms a transient-execution attack into a side-channel attack, leaking metadata instead of data. 
We show that despite software mitigations, attackers can still leak metadata from other security domains by observing the success rate of Meltdown on non-secret data.
With \IDTAttack, we present the first cache-line granular monitoring of kernel addresses.
\IDTAttack allows an attacker to obtain cycle-accurate timestamps for attacker-chosen interrupts.

We use our attack to get accurate inter-keystroke timings and fingerprint visited websites. 
While we propose a low-overhead software mitigation to prevent the exploitation of \IDTAttack, we emphasize that the side-channel aspect of transient-execution attacks should not be underestimated.
\end{abstract}

\section{Introduction}\label{sec:intro}

Microarchitectural side-channel attacks have been known for several years~\cite{Kocher1996}.
These attacks exploit the side effects of CPU implementations to infer metadata about processed data.
Well-known examples of microarchitectural side-channel attacks include cache attacks, \eg \FlushReload~\cite{Yarom2014Flush} or \PrimeProbe~\cite{Percival2005}, which have been used to leak cryptographic secrets~\cite{Aciicmez2008,Yarom2014Flush} or violate the privacy of users, \eg by spying on user input~\cite{Oren2015,Gruss2015Template,Schwarz2018KeyDrown}.
The discovery of transient-execution attacks, such as Meltdown~\cite{Lipp2018meltdown} and Spectre~\cite{Kocher2019}, was a game changer for microarchitectural attacks, as these directly leak data instead of metadata.
Hence, even best practices for side-channel-resistant software~\cite{Intel2020Guidelines,BSISidechannels1} do not protect secrets anymore. 
In Meltdown attacks, architecturally inaccessible data is accessed during out-of-order execution and encoded into a microarchitectural element, \eg the cache, protected from the pipeline flush~\cite{Lipp2018meltdown,Ragab2021rage}.
A subsequent side-channel attack, \eg \FlushReload, converts the microarchitectural into an architectural state, revealing the data.

As only new CPU generations contain hardware fixes for Meltdown-type attacks, short- and mid-term mitigations rely on software workarounds.
These workarounds ensure that no confidential data is stored in affected buffers when untrusted code is executed~\cite{VanSchaik2019RIDL,Schwarz2019ZL,IntelMDSadvisory} or that the victim data is not addressable~\cite{Gruss2018Kernel,Vanbulck2018foreshadow}.
For \meltdownus~\cite{Lipp2018meltdown}, \ie the original Meltdown attack, the OS unmaps the majority of its address space while running in user space, making sensitive data non-addressable~\cite{Gruss2017KASLR}.
The remaining mapped pages are not considered confidential, such that \meltdownus is considered not exploitable. 
On Linux, this technique is implemented as kernel page-table isolation (KPTI)~\cite{KaiserLKML2017kpti}.

In this paper, we show that even with state-of-the-art mitigations, Meltdown can be transformed from a transient-execution attack into a side-channel attack.
The main idea is based on two properties.
First, while KPTI unmaps most kernel pages, several kernel pages with non-secret content are necessary on x86 CPUs and cannot be unmapped in user space.
Second, Meltdown~\cite{Lipp2018meltdown} can only leak data if it is cached in the L1D cache, making it usable as a cache-state oracle.
Combining these two properties leaks the meta information on whether (non-confidential) kernel data was accessed.
Hence, Meltdown can be used as a high-resolution cache attack with cache-line granularity on the kernel.
This side channel is superior to state-of-the-art cache attacks on the kernel, which only achieve page~\cite{Lipp2022Prefetch} or cache-set granularity~\cite{Schwarz2018KeyDrown}.

\noindent We gain an interesting insight from this attack:

\textit{While a layer of indirection is necessary for Meltdown to leak data, another layer of indirection transforms the attack to leak metadata of architecturally inaccessible data.}

In other words, exploiting a modified version of the Meltdown attack enables the leakage of metadata that cannot be leaked in this granularity with a traditional side-channel attack.

Based on this, we present \IDTAttack, a side-channel attack able to spy on interrupts.
We exploit that the interrupt descriptor table (IDT) must always be mapped on x86~\cite{Intel_vol1,Gruss2017KASLR}.
Hence, despite software mitigations such as KPTI, an attacker can use the side channel to monitor interrupt activity.
In contrast to previous works that exploit interrupts as a side channel~\cite{Lipp2017Interrupt,Schwarz2018KeyDrown,Vanbulck2018nemesis}, \IDTAttack can target specific interrupts, \eg network or keyboard interrupts, instead of just observing that \emph{any} interrupt occurred and works for unprivileged attackers.
We identify which website a user visits from the Alexa top 15 and top 100 websites with a precision of \SI{80}{\percent} and \SI{55}{\percent}, respectively.
Furthermore, we reliably observe keystroke timings with an average F-score of \SIx{0.89}.
We propose to mitigate \IDTAttack by marking the IDT uncachable, preventing any entry from being cached.
This mitigation is practical, with an average performance overhead of less than \SI{0.5}{\percent} in 5 different benchmarks simulating real-world workloads. %

Our attacks show that while mitigating data leakage is essential, the side-channel aspect of such fixes can be overlooked.
We show that adding additional layers of indirection to existing attacks can change their properties.
As a result, we create a new side-channel attack from a CPU vulnerability commonly considered unexploitable when applying state-of-the-art software mitigations.
Hence, we argue that future software workarounds should consider the side-channel aspect to prevent such attack vectors.
Thus, we encourage researchers to look at other mitigations for hardware vulnerabilities to determine whether they can be circumvented to repurpose the underlying vulnerability as a side channel.
For this purpose and to ease reproducibility, we open-source the code of our findings on GitHub\footnote{\url{https://github.com/cispa/indirect-meltdown}}.

To summarize, we make the following contributions:
\begin{compactenum}
    \item We show that adding another layer of indirection to Meltdown transforms Meltdown into a side channel that infers the cache state of non-sensitive kernel pages with cache line granularity, leaking details about, \eg interrupts. %
    \item We use our side channel to detect the visited websites of a user and spy on their keystroke timings.
    \item We present a practical mitigation that stops our attack, while introducing an average overhead of less than \SI{0.5}{\percent} for real-world workloads.
\end{compactenum}

\textbf{Responsible Disclosure.}
We disclosed our findings to Intel on February 15, 2023 and AMD on February 16, 2023.
Despite both vendors acknowledging our findings, they informed us that they do not plan to roll out further mitigations.
\section{Background} \label{sec:background}
In this section, we provide the background for this paper. 
We introduce side channels, transient-execution attacks, and the interrupt descriptor table. 

\subsection{Side Channels}

Side channels leak metadata of (secret) information.
In a side-channel attack, an attacker infers secrets from this metadata.
For leaking metadata, secret-dependent observable differences must exist, \eg response time or power consumption that depends on the bits of a cryptographic key.
Previous research showed that side channels can be practical tools in an attacker's repertoire~\cite{Kocher1996,Monaco2018}, especially for attacking cryptographic implementation~\cite{Kocher1996,Monaco2018}.
In recent years, researchers have shown various side-channel attacks exploiting microarchitectural components~\cite{Monaco2018,Oren2015,Weber2021Osiris,Wang2022hertzbleed,Pessl2016}.
These components include CPU caches~\cite{Yarom2014Flush,Percival2005}, branch predictors~\cite{Bhattacharya2015,Aciicmez2007predicting}, execution units~\cite{Gras2020,Weber2021Osiris}, DRAM components~\cite{Pessl2016}, and power usage~\cite{Yan2015}. %
The fundamental property that enables such microarchitectural side-channel attacks is that different processes share many hardware components.
Hence, the resource usage of one process affects the possible resource usage of another process, leaking meta information between the processes.

\subsection{Transient-Execution Attacks}

Two important performance optimizations in modern CPUs are out-of-order execution and speculative execution.
Out-of-order execution allows the CPU to reorder or parallelize the execution of instructions in the instruction stream.
Speculative execution predicts the outcome of branch and memory load instructions, reducing pipeline stalls.
Executed instructions that never commit their state changes to the architecture due to a misspeculation or preceding fault are called transient instructions~\cite{Canella2019A,Intel2020Refined}.
Transient-execution attacks~\cite{Canella2019A} exploit transient instructions to read otherwise inaccessible memory~\cite{Lipp2018meltdown,Kocher2019}.
While transient instructions do not have an architectural effect, they can influence microarchitectural states, such as cache states.
These traces can be converted to architectural states using a microarchitectural side channel, \eg \FlushReload.
In recent years, researchers and CPU vendors discovered a variety of transient-execution attacks~\cite{Lipp2018meltdown,Kocher2019,Canella2019Fallout,Schwarz2019ZL,VanSchaik2019RIDL,Moghimi2020medusa,Vanbulck2020lvi,Maisuradze2018spectre5,Koruyeh2018spectre5,Schwarz2019STL,Ragab2021crosstalk,Ragab2021rage,Canella2019A}.

One category of transient-execution attacks are Meltdown-type attacks~\cite{Canella2019A}.
The first discovered Meltdown-type attack, later referred to as \meltdownus~\cite{Canella2019A}, allows unprivileged attackers to leak cached kernel memory. 
After a faulting load to a kernel address, the value is transiently available and can be encoded in the microarchitecture, \eg in the cache. 
The attacker decodes the encoded value using a side channel, \eg using \FlushReload.
Meltdown-type attacks, and especially \meltdownus, affect a variety of modern CPUs~\cite{Intel2023Affected}.

\subsection{Interrupt Descriptor Table (IDT)}%
\label{sec:idt}

Devices, such as network interface controllers or keyboards, use interrupts to notify the OS of events, \eg incoming network packets or key presses.
On an interrupt, the CPU switches to ring 0 and looks up the corresponding interrupt service routine (ISR) for the specific interrupt in the interrupt descriptor table (IDT).
The CPU interrupts the current execution and jumps to the ISR to handle the interrupt.
After handling the interrupt, the CPU continues executing the previous instruction stream. 
We only consider the 64-bit x86 IDT.
Each core can have its own IDT containing 256 interrupt vectors~\cite[Chapter 6.10 \& 6.14]{Intel_vol3}.
Each interrupt vector is 16 bytes in size and represents one device~\cite[Chapter 6.10 \& 6.14]{Intel_vol3}.
Hence, the IDT has a total size of \SI{4}{\kilo\byte}, \ie one memory page, and is stored in the main memory.
Each of these interrupt vectors essentially consists of a 64-bit (8-byte) pointer to its ISR in the kernel.
The remaining 8 bytes store additional meta-information about the interrupt, such as the type and the privilege level of the interrupt~\cite[Chapter 6.14]{Intel_vol3}.
The base pointer to the IDT is stored in a CPU-internal register, which can be read with the \instr{sidt} instruction.
On modern Linux systems, the IDT is hard-coded to \code{0xfffffe0000000000}~\cite{KernelMemoryMap}.

\section{Meltdown as a Side Channel} \label{sec:meltdownsc}

\begin{figure}[t]
    \centering
    \input{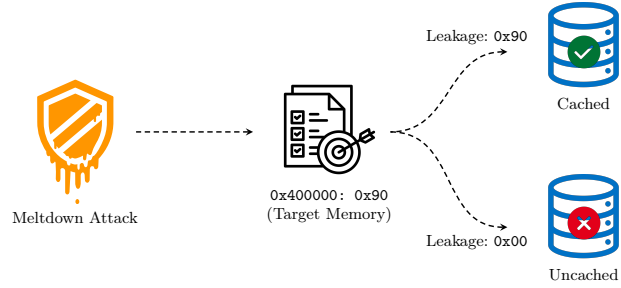}
    \caption{Meltdown as a side channel. The Meltdown attack only leaks data if the target address is in the L1D cache. Otherwise, the value \code{0x00} is leaked.}
    \label{fig:masc}
\end{figure}
In this section, we introduce the concept of transforming the transient-execution attack Meltdown into a side channel. 
The main idea is that the success rate of \meltdownus reveals the cache state of the target memory address.
We discuss which kernel memory ranges are still mapped despite the KPTI mitigation and how \meltdownus can be used to leak metadata about these memory pages.
For a list of CPUs affected by \meltdownus and thus affected by our attack, we refer the reader to the Intel's list of vulnerable CPUs~\cite{Intel2023Affected}.

While Lipp~\etal\cite{Lipp2018meltdown} discussed that \meltdownus works best if the target address is stored in the L1D cache, Xiao~\etal\cite{Xiao2019Speechminer} and Schwarzl~\etal\cite{Schwarzl2021Speculative} show that \meltdownus is limited to the L1D cache. 
Leakage from other cache levels is only caused by prefetching the data into the L1 cache, \eg via speculative execution. 
We exploit this requirement to use Meltdown as a side channel:
If data is leaked via \meltdownus, it is in the L1D cache. 
An illustration of this concept is given in \Cref{fig:masc}.

By detecting whether the target memory address can be leaked, we learn whether it was previously accessed.
If the cache-line content can be leaked, the cache line is cached in the L1D, which is only the case if the cache line was recently accessed. 
As this attack can be applied to any mapped memory address, we can also use it on kernel memory pages that are mapped while in userspace.
This converts \meltdownus into an \EvictReload-style side channel for kernel memory.

\begin{listing}[tb]
 \begin{lstlisting}[style=customasm]
; rax = kernel address, rcx/rbx= probe page 1/2, 
cmp     [rax], 0x0
cmovne   rcx, rbx
mov      rax, [rcx]
\end{lstlisting}
\caption{Using \meltdownus as a side channel. If the target kernel address is cached, the user address stored in \instr{RBX} is cached. Otherwise, the user address stored in \texttt{RCX} is cached.}
\label{lst:meltdown-sc}
\end{listing}

\parhead{Attack Details}
\Cref{lst:meltdown-sc} shows the implementation of the encoding step when using \meltdownus as a side channel.
We compare the content of the kernel address to zero (Line 2) and, based on the result, select (Line 3) and access (Line 4) one out of two different pages. 
This works as the access transiently results in a zero if no value can be leaked by \meltdownus.
Otherwise, the result is non-zero if the targeted memory address is non-zero.
This code sequence is simpler than the \meltdownus code~\cite{Lipp2018meltdown} that transiently loads the value at the kernel address into a register and accesses one out of 256 pages based on the loaded value, since we only need to consider two cases, \ie cached and non-cached.
This means that instead of monitoring 256 cache lines, our attack only has to monitor a single cache line.
In line with the \meltdownus attack, this code snippet raises an exception that has to be handled, \eg with fault handling, TSX, or fault suppression via speculation~\cite{Lipp2018meltdown}. 
For the decoding, \ie transferring the information encoded in the microarchitecture to an architectural state, any side channel can be used. 
For simplicity and in line with related work~\cite{Lipp2018meltdown,Vanbulck2018foreshadow,Schwarz2019ZL,VanSchaik2019RIDL,Canella2019Fallout}, we rely on \FlushReload to recover the encoded information.
In case of a recent access by the victim, the target address is stored in the L1D cache.
Thus, to monitor further cache accesses, we need to remove the target address from the L1D cache.
As the target memory address cannot be accessed, we need to rely on eviction.
However, as the L1D cache is virtually-indexed, evicting from it is straightforward and can be achieved by accessing virtual memory addresses falling into the same cache line as the target address. 
Note that we only need to evict the target address when an access occurred, as the Meltdown attack itself does not cache the target address.

\parhead{Attack Surface}
We investigate the attack surface of using \meltdownus as a side channel by analyzing which kernel pages are mapped in user space when KPTI is active. 
As \meltdownus cannot be fixed via microcode on affected hardware, KPTI~\cite{Gruss2018Kernel} is used as a software workaround on \meltdownus-affected CPUs.
KPTI ensures that while an application runs in user space, no kernel page containing sensitive information is mapped into the address space. 
For this, KPTI relies on a second set of page tables~\cite{Gruss2017KASLR}.
However, while this works theoretically, x86 always requires some kernel pages to be mapped, even when running in user space. 
Luckily, the content of these pages, \eg the IDT, can be chosen such that they do not contain secrets. 

We investigate which pages are still mapped in userspace by iterating through the user page tables using the kernel module PTEditor~\cite{SchwarzPteditor}.
For the user-page-table root, we set bit 11 of the physical address stored in the kernel \instr{CR3} register~\cite{Gruss2017KASLR}.
We iterate through the mappings in the upper half of the address space for kernel addresses mapped in user space.
We discover between 198 and 394 \SI{4}{\kilo\byte} kernel pages mapped in user space, depending on the CPU. 
However, these pages can be classified into only 3 distinct ranges. 
The first range is the kernel entry.
This range has been exploited for microarchitectural KASLR breaks~\cite{Schwarz2019STL,Canella2020kaslr,Weber2021Osiris}. 
The second range is used for descriptor tables, such as the interrupt-descriptor table or the global-descriptor table.
Finally, the third range is within the range of the direct physical map~\cite{KernelMemoryMap}, mapping 4 physical pages. 
One of these mappings is to the task state segment, which is also mapped directly. 
We cannot explain the reason for these remaining mappings, as the target is already mapped in user space. 
Still, this does at least not increase the attack surface. 
The most interesting target for using \meltdownus as a side channel is the IDT (\cf \Cref{sec:meltdownsc:idt}).

\section{\IDTAttack}\label{sec:meltdownsc:idt}
In this section, we introduce \IDTAttack, a side-channel attack that precisely detects when an attacker-chosen interrupt occurs. 
\IDTAttack achieves that by observing the cache state of the IDT entries of the targeted interrupts. 

Linux uses one IDT per core that always resides at the same location (\cf \Cref{sec:idt}).
This IDT is mapped in all processes, even with KPTI.
Hence, our attack can target the IDT despite applied software-based \meltdownus mitigations.
Note that a different operating system could randomize the location of the IDT upon booting and thus harden the system against our attack.

\begin{figure}[t]
    \centering
    \input{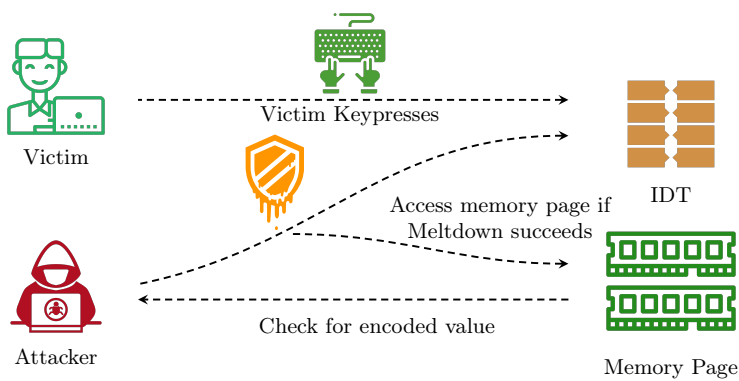}
    \caption{Using \IDTAttack to leak interrupts, such as keystrokes.}
    \label{fig:idtattack-overview}
\end{figure}

\parhead{Attack Overview}
\Cref{fig:idtattack-overview} shows an overview of \IDTAttack.
We use \meltdownus to read a specific IDT entry corresponding to a targeted interrupt. 
IDT entries are accessed---and thus cached in L1D---if the CPU core handles an interrupt.
Hence, if the leakage of the entry is successful, we infer that the interrupt was triggered; otherwise, it was not.
Consequently, with \IDTAttack we know the timestamp when the interrupt occurred. 
Note that due to the CPU's hardware prefetchers the actual accuracy of our attack is reduced to blocks of 8 adjacent IDT entries.
Further details on this are discussed later in this section.
When detecting an interrupt, \IDTAttack uses eviction to remove the targeted IDT entry from the L1D cache again. 
This is crucial for the attack as after every observed interrupt, the attacker must ensure that the IDT entry is removed from the cache as quickly as possible.
Otherwise, subsequent accesses to that memory address, \ie subsequent interrupts of the same type, cannot be detected.

\parhead{Threat Model}
Our attack requires a victim application that leaks information by having secret or data-dependent interrupts.
Such a victim can, \eg receive keystrokes~\cite{Gruss2015Template,Lipp2017Interrupt}, issue secret-dependant legacy syscalls~\cite{Zhang2009}, or communicate over the network~\cite{Zhang2023MWAIT}.
Besides this, we assume a bug-free software containing no logical vulnerabilities.
We further consider the attacker and victim both executing unprivileged native code on the same \meltdownus-affected CPU. %
The attack does not assume any disabled mitigations, \ie it works with state-of-the-art software-based Meltdown mitigations.

\parhead{Implementation}
For inferring the cache state of the IDT entry, we use the code from \Cref{lst:meltdown-sc}. 
Note that each IDT entry is 16 bytes in size. 
Thus, there are 4 IDT entries per cache line that are all cached when an interrupt occurs.
The exact offset of the IDT entry we are targeting with \IDTAttack is irrelevant, as every interrupt corresponding to that entry caches the entire cache line.
One should note that the granularity of our attack in a normal environment is restricted to blocks of 8 IDT entries.
The reason is that upon receiving an interrupt, the CPU's adjacent cache-line prefetcher puts two adjacent cache lines, \ie 8 adjacent IDT entries, into the L1D cache at once.

To detect the correct entry in the IDT, we template the IDT entries. 
First, we record the number of interrupts for every IDT entry over a fixed time window, \eg \SI{100}{\milli\second}. 
Second, we repeat this recording step while inducing the interrupt in parallel. 
Depending on the type of interrupt, this can be done in soft- or hardware. 
Some interrupts can be triggered the same way the victim triggers the interrupt, \eg sending a network packet for network interrupts. 
If this is not possible, \eg for keyboard interrupts, an attacker can induce the same interrupt as a software interrupt, using the \instr{int} instruction. %
If the difference in the number of interrupts correlates with the induced interrupts, the correct IDT entry is identified.
As we do not require fine-grained measurements for this step, we can take the information exposed by the Linux interface, \ie the file \code{/proc/interrupts}.

As discussed in \Cref{sec:meltdownsc}, to ensure that \IDTAttack can detect more than the first interrupt, the IDT entry has to be evicted again from the L1D cache.
The cache replacement policy on our machines is Tree-PLRU~\cite{Abel2019uops}, and the cache is virtually indexed using bits 6 to 11. 
Thus, we access memory addresses falling into the same L1D cache set by accessing pages at the same offset as IDT entry offset, which performs well enough for the attacks. 

\section{Evaluation} \label{sec:eval}

In this section, we evaluate the performance and reliability of \IDTAttack.
All experiments are executed on an \TSXLaptopCPU running \TSXLaptopOS with \TSXLaptopKernel.
On a general level, \IDTAttack allows observing the cache state of an inaccessible but mapped memory page.
More precisely, we can distinguish between a memory address that is cached in the L1D cache and a memory address that is not cached in the L1D. %

First, we evaluate how precisely we can distinguish between such two memory addresses.
We mount our exploit on two memory addresses, one being cached in the L1D cache and one not being cached.
Note that distinguishing between an address cached in L1D and not cached at all is enough for an attacker to mount side-channel attacks.
Our tests show that for a memory address cached in L1D, we have a successful leak in \SI{99.6}{\percent} of cases and no leakage in \SI{100}{\percent} of cases for uncached memory addresses.
We observe that for the uncached target byte, we only see the byte 0x0 encoded in our lookup array.
This observation is in line with previous work~\cite{Lipp2018meltdown,Xiao2019Speechminer}.
These results show that an attacker can reliably infer the cache state of the target kernel memory address by observing whether the \meltdownus leakage exists.

\begin{figure}[t]
    \centering
    \begin{tikzpicture}
\begin{axis}[
style={font=\footnotesize},
scaled y ticks=true,
width=\hsize,
height=3.0cm,
xmin=2000,
xmax=22000,
legend style={at={(0.25,1)}},
xlabel={Delay between interrupts [cycles]},
ylabel={\parbox{2cm}{\centering Observed \\interrupts [\%]}},
]
\addplot+[blue,thick,mark=none] table[x=interrupts,y=tsx,col sep=comma] {data/tsx_sig_accuracy.csv};

\addplot+[red,thick,mark=none] table[x=interrupts,y=tsx,col sep=comma] {data/tsx_sig_accuracy_pp.csv};

\end{axis}
\end{tikzpicture}
    \caption{Delay between interrupts and number of interrupts missed by \IDTAttack (upper line) and \PrimeProbe (lower line).}
    \label{fig:perfattack-interrupt-intervals}
\end{figure}
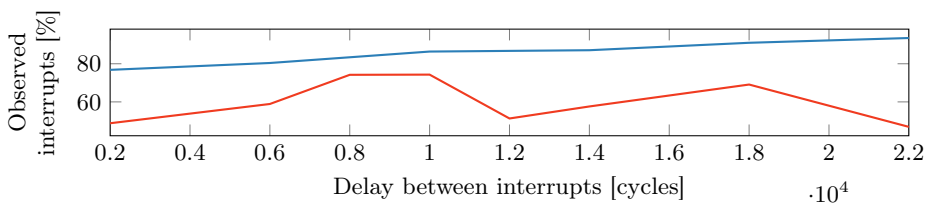
\Cref{fig:perfattack-interrupt-intervals} shows how different delays between interrupts interfere with the observation rate of our attack, \ie the number of interrupts successfully detected.
More precisely, we trigger \SIx{10000} interrupts with an artificial busy wait of $n$ cycles between them.
This allows us to measure the success rate of our attack when the victim triggers interrupts at a high frequency.
We observe that if the interrupts are more closely spaced than \SIx{25000} cycles, our detection rate decreases.
We further observe that for interrupts happening at a slower rate, we have success rates of up to \SI{99.5}{\percent}.
Thus, attackers can exploit \IDTAttack to reliably leak interrupts up until this frequency.

\parhead{Comparison to Related Kernel Attacks}
To the best of our knowledge, \IDTAttack is the first cache-line-granular side-channel attack on the kernel.
\IDTAttack does not require read- or writable shared memory, which is typical for cache attacks~\cite{Yarom2014Flush,Gruss2016Flush,Lipp2016}, preventing their use on kernel memory.
While there are also cache attacks not requiring shared memory~\cite{Briongos2020reload,Purnal2021PrimeScope,Disselkoen2017prime,Percival2005}, \IDTAttack yields a better granularity as it allows targeting specific cache lines of the kernel.
Additionally, cache attacks without shared memory often require knowledge of physical addresses to construct reliable and efficient eviction sets~\cite{Tromer2010}.
As we do not assume that knowledge in our threat model, we compare \IDTAttack with \PrimeProbe on the L1D, as this attack has the same threat model.

Not only does \IDTAttack have a finer granularity, but it also outperforms \PrimeProbe in terms of reliability.
\Cref{fig:perfattack-interrupt-intervals} shows the number of interrupts missed by our \PrimeProbe implementation.
Note that our implementation only counts an interrupt if two probe steps show higher access timing.
While this may not be optimal, it significantly reduces the number of false positives and shows the best performance during our evaluation.
We suspect that the reason for this is that the probes execute fast enough to measure the activity on the IDT entry multiple times during the interrupt handling.
To further compare the two side channels, we compare their performance in a more artifical scenario.
We take \SIx{100000} measurements for each attack while the victim accesses the targeted cache line \SIx{50000} times per attack.
Finally, we compare the results of our side-channel attacks to the ground truth of victim accesses.
For \IDTAttack, we get a recall of \SIx{0.999} and a precision of \SIx{1.0}, yielding an F-score of \SIx{0.999}.
For \PrimeProbe on the L1D, we measure a recall of \SIx{1.0} and a precision of \SIx{0.834}, yielding an F-score of \SIx{0.91}.

\section{Case Studies} \label{sec:case}
In this section, we introduce 2 case studies demonstrating \IDTAttack.
Leveraging \IDTAttack, we show that an attacker can spy on websites visited by a victim on the same system (\cf\Cref{subsec:website-fp}).
Furthermore, we show that fine-grained timing measurements of interrupts leak information about the keystrokes entered by a user (\cf\Cref{subsec:keystrokes}).

\subsection{Website Fingerprinting}\label{subsec:website-fp}

In this section, we use \IDTAttack to detect which website a user opens by monitoring network interrupts.
For this purpose, we perform the website fingerprinting attack on an \LabTwelveCPU, with \LabTwelveOS and \LabTwelveKernel.

\parhead{Threat Model}
In line with previous work~\cite{Zhang2009,Jana2012,Spreitzer2016fingerprint,Lee2014stealing,Gulmezoglu2017perfweb}, we assume an unprivileged attacker with native code execution on the victim system.
In contrast to these works, we do not rely on OS interfaces, as they are nowadays only available to privileged users. %
We assume the attacker application runs on the physical core that handles the network interrupts, which the unprivileged \code{pthread\_setaffinity\_np} Linux API can achieve.

\parhead{Attack Overview}
We do not assume prior knowledge of the IDT entry that the attacker needs to probe.
Thus, the first step of the attack is to find the specific IDT entry that handles the network interrupts.
To do that, we use \IDTAttack on all IDT entries while introducing additional network traffic.
For each entry, we record the number of accesses during a short fixed period, \eg \SI{1}{second}.
Next, we repeat the measurement without generating additional network interrupts.
As the network interrupts bring the specific IDT entries into the cache, entries with the most significant differences in the number of accesses are likely related to the network interrupts.

In line with previous work~\cite{Zhang2023MWAIT}, we rely on a coarse-grained timer, \eg \texttt{clock\_gettime} or \texttt{setitimer}, to record the number of interrupts per \SI{5}{\milli\second} interval when a user opens a website.
We then train a random forest classifier to fingerprint the opened website.

\begin{figure}[t]
    \centering
    \resizebox{0.9\hsize}{!}{
    \input{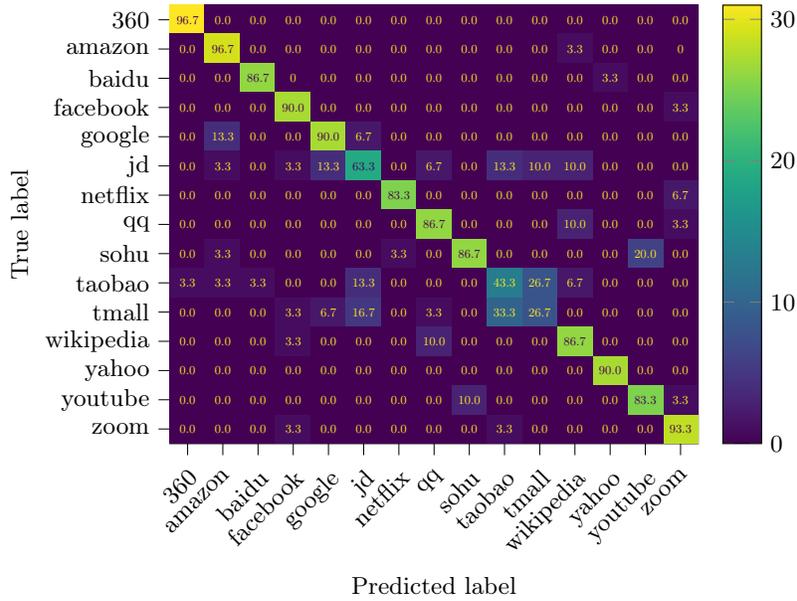}
    }
    \caption{The confusion matrix for the website classification. Given the Alexa top \num{15} websites, the trace is classified correctly with an overall probability of \SI{80}{\percent}.}
    \label{fig:website-fp}
\end{figure}

\parhead{Results}
We collect \num{100} interrupt traces for each of the Alexa \num{100} most-visited websites.
Each trace collects the number of interrupts in a \SI{5}{\milli\second} interval \num{400} times (\SI{2}{\second} in total).
The dataset is split into a training set of \SIx{7000} and a test set of \SIx{3000} examples, and the \code{n\_estimators} for the random forest classification are set to the default of \num{100}.
For the top \num{15} websites, we achieve a precision of \SI{80}{\percent} and a recall of \SI{81}{\percent}, as illustrated in the confusion matrix in \Cref{fig:website-fp}.
For the top \num{100} websites, we achieve an overall precision of \SI{55}{\percent} and a recall of \SI{56}{\percent}.
Note that a more precise timer, \ie with a better accuracy than \SI{5}{\milli\second} would likely improve these results.

\parhead{Comparison to Related Work}
While Spreitzer~\etal\cite{Spreitzer2016fingerprint} report \SI{89}{\percent} accuracy on \num{100} sites on Android, the attack requires the unprivileged interface for sampling data-usage statistics.
Zhang~\etal\cite{Zhang2023MWAIT} report \SI{71}{\percent} accuracy on \num{100} sites on Intel, relying on the new \instr{umwait} instructions only available on the latest Intel microarchitectures.
The interrupt attack by Lipp~\etal\cite{Lipp2017Interrupt} correctly classifies a website in \SI{81.75}{\percent} of cases inside the browser when only looking at \num{10} websites.
Lee~\etal\cite{Lee2014stealing} exploit GPU vulnerabilities and report \SI{69.4}{\percent} and \SI{60.9}{\percent} with two different techniques on \num{100} sites randomly chosen from Alexa Top \num{1000}.

\subsection{Keystroke Timings via \IDTAttack}\label{subsec:keystrokes}
In this section, we show that \IDTAttack can be used for keystroke-timing attacks, as first discussed by Song~\etal\cite{Song2001}.
We show that \IDTAttack reliably recovers keystroke timings on USB keyboards on an \LabTwelveCPU, with \LabTwelveOS and \LabTwelveKernel.

\parhead{Threat Model}
We assume an unprivileged attacker with native code execution on a system vulnerable to \IDTAttack. %
We further assume that the attacker application can be pinned to specific physical cores by unprivileged APIs.

\parhead{Experiment Setup}
In line with the first case study, we do not assume knowledge of the IDT entry. 
Thus, an attacker trying to locate the core responsible for handling keyboard interrupts can probe all interrupts on all cores for a short and fixed time interval.
Afterward, when the attacker knows that the victim is likely pressing keys, \eg by checking for interactive applications in the list of running processes, the attacker can probe these interrupts again and check for significant differences.
To optimize the measurements for this case study, the attacker pins the spy process on the sibling of the previously identified core.

We perform our experiments in two settings.
In the first setting, a lab environment, the eXtensible Host Controller Interface (xHCI) interrupts are handled by an isolated core.
In the second setting, a realistic environment, we boot the system without any preparations and simulate heavy system load with the stress utility (\code{stress -m 2 -c 2}).
The kernel distributes the interrupts over the available 4 cores.
xHCI interrupts share their core only with peripheral network interrupts in our experiments.
These interrupts occur every \SI{2}{\second}.

We spawn two processes.
The first one reads characters from \code{stdin} and logs microsecond timestamps of the keystrokes.
This process can be spawned on any core and provides ground-truth data.
The second process is pinned to the physical core handling xHCI interrupts.
This process logs microsecond timestamps of leaked interrupts via \IDTAttack.

We perform 3 runs of typing 200 random keys on the keyboard for both setups.
In our case study, all inputs are entered by a single person.
We record the timestamp traces of both processes.
We then match every recorded interrupt timestamp to the nearest ground truth timestamp.
Since xHCIs generate two interrupts for USB keyboards, \ie key down and key up, we assume two captured interrupts per actual timestamp.
Even though the difference between key down and key up events can improve the results of keystroke attacks~\cite{Pinet2016}, we choose to ignore their impact in this case study to focus on the concept.
Any missing timestamp from the expected 2 interrupts for each actual timestamp is counted as a false negative.
Any detected interrupt matching with more than one uniquely identifiable key-up and key-down event is counted as a false positive.

\begin{table}[tb]
\centering
\caption{Results for the inter-keystroke timing attack.}
\adjustbox{max width=\hsize}{
\begin{tabular}{cccccr}
    \toprule
    \textbf{Run}& \textbf{Noise}& \textbf{Recall}& \textbf{Precision}& \textbf{F-score}& \textbf{Delay (std dev.)}\\
    \midrule
    1& no& 0.93& 0.89& 0.91& \SI{-323}{\micro\second}  (\SI{35.66}{\micro\second})\\
    2& no& 0.91& 0.95& 0.93& \SI{-334.5}{\micro\second}  (\SI{29.71}{\micro\second})\\
    3& no& 0.90& 0.90& 0.90& \SI{-324}{\micro\second}  (\SI{34.11}{\micro\second})\\
    1& yes& 0.89& 0.87& 0.88& \SI{-573}{\micro\second} (\SI{64.25}{\micro\second})\\
    2& yes& 0.88& 0.88& 0.88& \SI{-568}{\micro\second} (\SI{49.46}{\micro\second})\\
    3& yes& 0.86& 0.86& 0.86& \SI{-551}{\micro\second} (\SI{56.28}{\micro\second})\\
    \bottomrule
\end{tabular}
}
\label{tab:keystroke}
\end{table}

\parhead{Results}
\Cref{tab:keystroke} shows the results for the 3 runs for both setups.
We calculate recall, precision, and F-score with the data acquired from matching recorded interrupts to ground truth timestamps.
We measure the median and the standard deviation of the delay when we detect the interrupts, showing that we detect keystroke interrupts around half a microsecond before they can be read from \code{stdin} in the victim application.
As expected, \IDTAttack performs slightly worse in the realistic setup compared to the isolated lab setup.
In the isolated setup, we observe an F-score of \SIx{0.91}, and for the realistic setup an F-score of \SIx{0.87}.
In comparison, the Android-based keystroke timing attacks from Schwarz~\etal\cite{Schwarz2018KeyDrown} achieve an F-score of \SIx{0.94} and \SIx{0.81}.
Similar attacks from Vila~\etal\cite{Vila2017} and Wang~\etal\cite{Wang2015mole} achieve a recall of \SIx{0.98} and \SIx{0.57}, respectively.
Thus, our results are comparable to previous work.
Note that depending on the goal of an attacker, further steps are required for an end-to-end attack, such as machine-learning-based password recovery or user classification.

\section{Mitigations} \label{sec:countermeasures}
In this section, we propose a mitigation against \IDTAttack.
We evaluate the mitigation and show that it only introduces a minimal performance overhead.

Although the root cause of \IDTAttack cannot be mitigated in software, we propose a software mitigation to prevent exploitation. %
The main idea is to ensure that the cache state of an IDT entry cannot be inferred by marking the IDT as uncachable, ensuring that the cache state is always the same.

\parhead{Implementation}
Linux uses a shared IDT across all CPU cores.
This single IDT is allocated once by the OS and keeps its physical location until reboot.
We rely on memory-type range registers (MTRRs) to mark the physical range of the IDT as uncachable.
While the number of MTRRs is limited~\cite[Chapter 11.11]{Intel_vol3}, we only require a single MTRR due to the shared IDT.
MTRRs have the advantage that the memory type defined by them cannot be overwritten. %

Alternatively, if no MTRR can be spared, the IDT mapping can be marked as uncachable via the memory type in the corresponding page-table entry.
Care has to be taken that this is done in every single user-space process, as well as in the kernel.
This requires more changes to the kernel and introduces a startup overhead for every application. 
Thus, we opted for the MTRR-based approach, requiring only a minimal overhead at boot for the configuration and allowing the implementation as a kernel module.

\parhead{Evaluation}
We evaluate the security and performance of our approach.
All evaluations are run on an Intel Xeon E3-1505M v5, with Ubuntu 20.04.1 and kernel 5.4.0.
For the security evaluation, we mount \IDTAttack with our active mitigation.
As expected, we do not see any leakage.
With the uncachable IDT, \IDTAttack can never leak an entry of the IDT, preventing \IDTAttack.

\begin{table}[tb]
  \caption{Performance of uncachable IDT on the SPEC CPU 2017 benchmark.}
  \begin{tabular*}{\hsize}{l@{\extracolsep{\fill}} rr r}
    \toprule
    \multirow{2}{*}{\textbf{Benchmark}} & \multicolumn{2}{c}{\textbf{SPEC Score}} & \multicolumn{1}{c}{\textbf{Overhead}} \\
    & Baseline & Uncachable & [\%]      \\
    \midrule
    600.perlbench\_s  & $1.88$ & $1.88$ & $~0.00\,\%$ \\
    602.gcc\_s        & $1.11$ & $1.11$ & $~0.00\,\%$ \\
    605.mcf\_s        & $1.91$ & $1.91$ & $~0.00\,\%$ \\
    620.omnetpp\_s    & $1.50$ & $1.52$ & $+1.33\,\%$ \\
    623.xalancbmk\_s  & $1.55$ & $1.58$ & $+1.94\,\%$ \\
    625.x264\_s       & $1.54$ & $1.54$ & $~0.00\,\%$ \\
    631.deepsjeng\_s  & $1.33$ & $1.33$ & $~0.00\,\%$ \\
    641.leela\_s      & $1.20$ & $1.20$ & $~0.00\,\%$ \\
    648.exchange2\_s  & $3.50$ & $3.52$ & $+0.57\,\%$ \\
    657.xz\_s         & $0.91$ & $0.91$ & $~0.00\,\%$ \\
    \midrule
    \textbf{Average}  &        &        & $+0.65\,\%$ \\
    \bottomrule
  \end{tabular*}
  \label{tab:speccpu_idt}
\end{table}

\begin{table}[tb]
    \caption{Kraken benchmark results.}
    \label{tab:kraken-benchmark}
    \centering
    \adjustbox{max width=\hsize}{
        \begin{tabular}{lrrr}
            \toprule
            \rotatebox{0}{\textbf{Test Case}} & \rotatebox{0}{\textbf{Baseline}} & \rotatebox{0}{\textbf{Uncacheable IDT}} & \rotatebox{0}{\textbf{Overhead}}\\ \midrule
            ai                       & \SI{164.8}{\milli\second} (+/- \phantom{1}\SI{6.0}{\percent})  & \SI{169.6}{\milli\second} (+/- \SI{6.0}{\percent})    & $+2.91\,\%$  \\
            astar                    & \SI{164.8}{\milli\second} (+/- \phantom{1}\SI{6.0}{\percent})  & \SI{169.6}{\milli\second} (+/- \SI{6.0}{\percent})    & $+2.91\,\%$  \\
            audio                    & \SI{532.8}{\milli\second} (+/- \phantom{1}\SI{2.6}{\percent})  & \SI{540.8}{\milli\second} (+/- \SI{2.1}{\percent})    & $+1.50\,\%$  \\
            beat-detection           & \SI{141.2}{\milli\second} (+/- \phantom{1}\SI{3.6}{\percent})  & \SI{141.7}{\milli\second} (+/- \SI{3.0}{\percent})    & $+0.28\,\%$  \\
            dft                      & \SI{115.8}{\milli\second} (+/- \phantom{1}\SI{3.7}{\percent})  & \SI{117.9}{\milli\second} (+/- \SI{5.2}{\percent})    & $+1.81\,\%$  \\
            fft                      & \SI{125.2}{\milli\second} (+/- \phantom{1}\SI{2.9}{\percent})  & \SI{125.8}{\milli\second} (+/- \SI{3.3}{\percent})    & $+0.48\,\%$  \\
            oscillator               & \SI{150.6}{\milli\second} (+/- \phantom{1}\SI{6.6}{\percent})  & \SI{155.5}{\milli\second} (+/- \SI{5.3}{\percent})    & $+3.25\,\%$  \\
            imaging                  & \SI{406.2}{\milli\second} (+/- \phantom{1}\SI{2.1}{\percent})  & \SI{400.3}{\milli\second} (+/- \SI{2.4}{\percent})    & $-1.45\,\%$  \\
            gaussian-blur            & \SI{158.3}{\milli\second} (+/- \phantom{1}\SI{4.0}{\percent})  & \SI{154.0}{\milli\second} (+/- \SI{2.3}{\percent})    & $-2.72\,\%$  \\
            darkroom                 & \SI{ 80.1}{\milli\second} (+/- \phantom{1}\SI{0.8}{\percent})  & \SI{79.6}{\milli\second} (+/- \SI{0.9}{\percent})     & $-0.62\,\%$ \\
            desaturate               & \SI{167.8}{\milli\second} (+/- \phantom{1}\SI{4.4}{\percent})  & \SI{166.7}{\milli\second} (+/- \SI{5.9}{\percent})    & $-0.66\,\%$  \\
            json                     & \SI{ 73.9}{\milli\second} (+/- \phantom{1}\SI{7.0}{\percent})  & \SI{76.7}{\milli\second} (+/- \SI{5.0}{\percent})     & $+3.79\,\%$ \\
            parse-financial          & \SI{ 37.0}{\milli\second} (+/- \SI{13.7}{\percent}) & \SI{37.7}{\milli\second} (+/- \SI{9.6}{\percent})     & $+1.89\,\%$\\
            stringify-tinderbox      & \SI{ 36.9}{\milli\second} (+/- \phantom{1}\SI{2.7}{\percent})  & \SI{39.0}{\milli\second} (+/- \SI{4.5}{\percent})     & $+5.69\,\%$ \\
            stanford                 & \SI{288.3}{\milli\second} (+/- \phantom{1}\SI{2.3}{\percent})  & \SI{287.0}{\milli\second} (+/- \SI{1.6}{\percent})    & $-0.45\,\%$  \\
            crypto-aes               & \SI{ 73.9}{\milli\second} (+/- \phantom{1}\SI{3.2}{\percent})  & \SI{74.1}{\milli\second} (+/- \SI{3.2}{\percent})     & $+0.27\,\%$ \\
            crypto-ccm               & \SI{ 65.6}{\milli\second} (+/- \phantom{1}\SI{4.1}{\percent})  & \SI{63.6}{\milli\second} (+/- \SI{2.4}{\percent})     & $-3.05\,\%$ \\
            crypto-pbkdf2            & \SI{100.0}{\milli\second} (+/- \phantom{1}\SI{1.9}{\percent})  & \SI{101.1}{\milli\second} (+/- \SI{1.7}{\percent})    & $+1.10\,\%$  \\
            crypto-sha256-iterative  & \SI{ 48.8}{\milli\second} (+/- \phantom{1}\SI{5.6}{\percent})  & \SI{48.2}{\milli\second} (+/- \SI{3.0}{\percent})     & $-1.23\,\%$ \\ \midrule
            \textbf{Total}           & \SI{1466.0}{\milli\second} (+/- \phantom{1}\SI{0.9}{\percent})  & \SI{1474.4}{\milli\second} (+/- \SI{0.7}{\percent})  & $+0.57\,\%$    \\
            \bottomrule                                                                         
        \end{tabular}                                                                          
    }
\end{table}

To evaluate the overhead of our mitigation, we execute benchmarks generating both high CPU loads and a large number of interrupts. 
We execute SPEC CPU 2017, which resembles generic real-world workloads, and additionally, Kraken and JetStream, two JavaScript benchmarks, to see the impact on web services.
For the baseline, we run the benchmarks on the unmodified system.
As marking the IDT as uncachable is implemented as a kernel module, we can run the benchmark on precisely the same kernel without even rebooting.
Hence, with this setup, there should not be any other factors influencing the benchmark results.
\Cref{tab:speccpu_idt} shows the results of the SPEC CPU benchmark.
The details of the JavaScript benchmarks can be found in \Cref{sec:appendix:js-benchmark}.
On average, we only measure a minimal performance overhead of \SI{0.65}{\percent} with SPEC CPU 2017, \SI{0.57}{\percent} with Kraken (\cf \Cref{tab:kraken-benchmark}), and \SI{0.32}{\percent} with JetStream.
We further test the impact on two interrupt-heavy benchmarks.
We execute the YCSB benchmark~\cite{Cooper2010YCSB} to evaluate the overhead for databases.
We test against a MongoDB instance and configure YCSB for \SIx{4500000} operations.
We observe an increase in interrupts of \SI{2326.29}{\percent}, \ie \SIx{52853.62} interrupts (on average over the 8 cores of the system), compared to the system idling for the same amount of time, \ie \SIx{2326.29} interrupts. 
As these benchmarks have a shorter execution time than the previous ones, we repeat this measurement 10 times on the baseline system and 10 times on the same system with the applied mitigation, thus ensuring a stable result.
We observe a median runtime of \SI{155155}{\milli\second} with a standard deviation of \SIx{243.01} for the baseline system and a median runtime of \SI{155181.5}{\milli\second} with a standard deviation of \SIx{286.99}, \ie an overhead of \SI{0.02}{\percent}.
To test the performance of a network-based key-value store, we evaluate the impact on a Memcached instance using the benchmarking framework mutilate~\cite{Leverich2014mutilate}.
We configure mutilate to execute 16 connections spanned over 8 threads.
\Cref{tab:mutilate-benchmark} shows the results.
Hereby, we observe an increase in interrupts of \SI{2630.30}{\percent}, \ie \SIx{69892.38} interrupts (on average over the 8 cores of the system), compared to the system idling for the same amount of time, \ie \SIx{2559.88} interrupts.
We execute this benchmark 10 times with and without the mitigation applied.
We observe a slowdown of the receive rate, the transmission rate, and the QPS of \SI{0.14}{\percent} each.

\begin{table}[tb]
    \caption{Mutilate benchmark results.}
    \label{tab:mutilate-benchmark}
    \centering
    \adjustbox{max width=\hsize}{
        \begin{tabular}{lrrr}
            \toprule
            \textbf{Attribute} & \textbf{Baseline score}              & \textbf{UC IDT score} & \textbf{Slowdown}\\
            \midrule
            QPS                & \SIx{176238.35} (std: \SIx{1376.41}) & \SIx{175987.5} (std: \SIx{1407.86}) & \SI{0.14}{\percent} \\
            RX                 & \SI{7759664293}{\byte} (std: \SIx{60596022.33}) & \SI{7748550743.5}{\byte} (std: \SIx{62033777.92}) & \SI{0.14}{\percent} \\
            TX                 & \SI{1208593212}{\byte} (std: \SIx{9439659.48}) & \SI{1206927213}{\byte} (std: \SIx{9614344.61}) & \SI{0.14}{\percent} \\
            \bottomrule                                                                         
        \end{tabular}                                                                           
    }
\end{table}

\section{Discussion} \label{sec:discussion}
In this section, we discuss Meltdown mitigations, their remaining leakage, and their applicability to other Meltdown variants, OS, and architectures.

\parhead{Meltdown Mitigations}
Gruss~\etal\cite{Gruss2017KASLR} showed that unmapping the kernel when possible mitigates several side-channel attacks on it.
This has become the state-of-the-art mitigation against \meltdownus~\cite{Lipp2018meltdown,Gruss2018Kernel}.
However, a limitation of the x86 architecture is that specific kernel structures, such as the IDT, must always be mapped.
While related work used these mappings to break KASLR~\cite{Schwarz2019STL,Canella2020kaslr,Weber2021Osiris}, such attacks can be prevented by using a different randomization offset for the pages that remain mapped.
However, this would not prevent \IDTAttack.
The reason is that \IDTAttack exploits the metadata of the data stored on kernel pages and not the content~\cite{Lipp2018meltdown} or the location~\cite{Schwarz2019STL,Canella2020kaslr,Weber2021Osiris}.

We show that uncachable memory eliminates the remaining leakage of KPTI.
Restricting the uncachable memory to the IDT ensures that the performance impact is minimal.
Hence, combining two incomplete mitigations for orthogonal problems hardens a system against side-channel attacks.

Ideally, vulnerabilities are mitigated in the hardware.
Still, despite hardware fixes, Canella~\etal\cite{Canella2020kaslr} showed that they leak metadata about the mapping of a virtual address.
While the leakage is much more limited than in our attack, it also shows that side-channel leakage can be overlooked when designing mitigations.

\parhead{Applicability to other Meltdown-Type Attacks}
Our attack is not limited to attacking the kernel.
While we convert \meltdownus into a side channel, the same technique can also be applied to other Meltdown variants.
For example, on CPUs affected by Foreshadow~\cite{Vanbulck2018foreshadow}, our technique could be used to implement an \EvictReload-style attack on Intel SGX enclaves.
For this, only the Meltdown attack has to be replaced with the related Foreshadow attack.
However, in contrast to the \meltdownus mitigations in the OS, the Foreshadow mitigations for SGX entirely prevent Foreshadow.
Hence, an enclave that can be attacked with Foreshadow as a side channel could also be attacked directly with Foreshadow. %
We leave it to future work to investigate whether other Meltdown-type attacks, such as RIDL~\cite{VanSchaik2019RIDL}, ZombieLoad~\cite{Schwarz2019ZL}, or Fallout~\cite{Canella2019Fallout}, could also be transformed into practical side-channel attacks. 

\parhead{Other OS and Architectures}
The underlying effects exploited in this paper are OS-agnostic.
While this paper targets Linux, we do not require any Linux-specific functionality.
For example, while the interrupt numbers differ on Windows, the mechanism is still the same.
The IDT is also mapped, as this is required by the x86 architecture, enabling \IDTAttack.

As \IDTAttack fundamentally relies on the \meltdownus CPU vulnerability, it does not apply to Meltdown-unaffected CPUs.
Hence, AMD and most Arm CPUs are not affected~\cite{Lipp2018meltdown}.
While there are Arm CPUs affected by \meltdownus~\cite{Lipp2018meltdown}, the interrupt handling is different, which would require adapting \IDTAttack to work with the IDT-equivalent, the Interrupt Vector Table (IVT).

\section{Conclusion} \label{sec:conclusion}
We showed that Meltdown cannot only act as a transient-execution attack but can also be exploited as a side-channel attack by adding another layer of indirection, despite active software mitigations.
We presented \IDTAttack, a side-channel attack that allows an attacker to monitor mapped kernel pages with cache-line granularity, enabling attackers to spy on chosen interrupts.
We showed that attackers can exploit this primitive to spy on websites visited by a user.
We analyzed that this fine-granular information leakage also reveals valuable insights into the typing behavior of a user by allowing to spy on their keystroke timings.
Hence, we conclude that even though \meltdownus is considered no longer exploitable, it still threatens the security of modern systems.
\section*{Acknowledgment}
We want to thank our anonymous reviewers for their comments and suggestions.
We also want to thank Leon Trampert and Niklas Flentje for providing their help with running the experiments. 
This work was partly supported by the Semiconductor Research Corporation (SRC) Hardware Security Program (HWS).

\bibliographystyle{IEEEtranS}
\bibliography{main}
\FloatBarrier

\appendix
\section{JavaScript Benchmark Results}\label{sec:appendix:js-benchmark}
\Cref{tab:jetstream-benchmark} shows the impact of our mitigations measured using the JetStream JavaScript benchmark.
The total overhead is \SI{0.23}{\percent}.
\begin{table}[!htb]
    \caption{JetStream benchmark results.}
    \label{tab:jetstream-benchmark}
    \centering
        \adjustbox{max width=0.9\hsize}{
        \begin{tabular}{lrrr}
            \toprule
            \textbf{Test Case} & \textbf{Baseline} & \textbf{UC IDT} & \textbf{Overhead}\\
            \midrule
            3d-cube-SP & 142.229 & 142.101 & $-0.09\,\%$ \\
            3d-raytrace-SP & 120.536 & 130.642 & $+8.38\,\%$ \\
            acorn-wtb & 14.267 & 13.714 & $-3.88\,\%$ \\
            ai-astar & 335.52 & 330.153 & $-1.60\,\%$ \\
            Air & 186.75 & 195.861 & $+4.88\,\%$ \\
            async-fs & 73.995 & 68.635 & $-7.24\,\%$ \\
            Babylon & 180.118 & 169.655 & $-5.81\,\%$ \\
            babylon-wtb & 14.468 & 16.173 & $+11.78\,\%$ \\
            base64-SP & 218.656 & 236.194 & $+8.02\,\%$ \\
            Basic & 195.428 & 168.823 & $-13.61\,\%$ \\
            bomb-workers & 20.826 & 17.721 & $-14.91\,\%$ \\
            Box2D & 120.487 & 144.29 & $+19.76\,\%$ \\
            cdjs & 32.632 & 28.819 & $-11.68\,\%$ \\
            chai-wtb & 42.27 & 42.068 & $-0.48\,\%$ \\
            coﬀeescript-wtb & 21.964 & 18.937 & $-13.78\,\%$ \\
            crypto & 279.665 & 341.319 & $+22.05\,\%$ \\
            crypto-aes-SP & 179.095 & 146.226 & $-18.35\,\%$ \\
            crypto-md5-SP & 113.45 & 106.569 & $-6.07\,\%$ \\
            delta-blue & 226.869 & 176.9 & $-22.03\,\%$ \\
            earley-boyer & 199.003 & 201.622 & $+1.32\,\%$ \\
            espree-wtb & 15.742 & 14.141 & $-10.17\,\%$ \\
            ﬁrst-inspector-code-load & 102.469 & 99.755 & $-2.65\,\%$ \\
            FlightPlanner & 176.477 & 176.196 & $-0.16\,\%$ \\
            ﬂoat-mm.c & 7.474 & 7.497 & $+0.31\,\%$ \\
            gaussian-blur & 225.208 & 230.45 & $+2.33\,\%$ \\
            gbemu & 62.156 & 57.95 & $-6.77\,\%$ \\
            gcc-loops-wasm & 21.568 & 22.449 & $+4.08\,\%$ \\
            hash-map & 94.658 & 124.756 & $+31.80\,\%$ \\
            HashSet-wasm & 27.422 & 29.125 & $+6.21\,\%$ \\
            jshint-wtb & 21.484 & 20.658 & $-3.84\,\%$ \\
            json-parse-inspector & 111.78 & 108.445 & $-2.98\,\%$ \\
        \end{tabular}
        \begin{tabular}{lrrr}
            json-stringify-inspector & 122.419 & 120.712 & $-1.39\,\%$ \\
            lebab-wtb & 24.557 & 24.599 & $+0.17\,\%$ \\
            mandreel & 32.562 & 32.546 & $-0.05\,\%$ \\
            ML & 13.853 & 13.274 & $-4.18\,\%$ \\
            multi-inspector-code-load & 109.236 & 92.512 & $-15.31\,\%$ \\
            n-body-SP & 466.978 & 459.006 & $-1.71\,\%$ \\
            navier-stokes & 400.438 & 409.595 & $+2.29\,\%$ \\
            octane-code-load & 502.996 & 460.544 & $-8.44\,\%$ \\
            octane-zlib & 14.938 & 15.063 & $+0.84\,\%$ \\
            OﬄineAssembler & 36.527 & 33.54 & $-8.18\,\%$ \\
            pdfjs & 75.934 & 78.712 & $+3.66\,\%$ \\
            prepack-wtb & 20.974 & 20.541 & $-2.06\,\%$ \\
            quicksort-wasm & 215.166 & 217.597 & $+1.13\,\%$ \\
            raytrace & 202.931 & 222.117 & $+9.45\,\%$ \\
            regex-dna-SP & 255.332 & 249.183 & $-2.41\,\%$ \\
            regexp & 281.028 & 279.361 & $-0.59\,\%$ \\
            richards & 196.298 & 189.976 & $-3.22\,\%$ \\
            richards-wasm & 37.949 & 33.478 & $-11.78\,\%$ \\
            segmentation & 11.835 & 12.609 & $+6.54\,\%$ \\
            splay & 88.279 & 85.402 & $-3.26\,\%$ \\
            stanford-crypto-aes & 173.872 & 188.774 & $+8.57\,\%$ \\
            stanford-crypto-pbkdf2 & 213.472 & 258.864 & $+21.26\,\%$ \\
            stanford-crypto-sha256 & 322.22 & 317.002 & $-1.62\,\%$ \\
            string-unpack-code-SP & 168.69 & 141.399 & $-16.18\,\%$ \\
            tagcloud-SP & 77.685 & 99.776 & $+28.44\,\%$ \\
            tsf-wasm & 42.163 & 67.481 & $+60.05\,\%$ \\
            typescript & 6.67 & 6.598 & $-1.08\,\%$ \\
            uglify-js-wtb & 12.796 & 13.636 & $+6.56\,\%$ \\
            UniPoker & 196.529 & 195.398 & $-0.58\,\%$ \\
            WSL & 0.411 & 0.405 & $-1.46\,\%$ \\ \midrule
            \textbf{Total} & 7909.404 & 7927.544 & $+0.23\,\%$    \\
            \bottomrule                                                                         
        \end{tabular}                                                                           
    }
\end{table}

\end{document}